\def\half{{1\over 2}}
\def\({\left (}
\def\){\right)}
\def\[{\left [}
\def\]{\right]}
\def\sqr#1#2{{\vcenter{\hrule height.#2pt\hbox{\vrule width.#2pt
height#1pt \kern#1pt \vrule width.#2pt}\hrule height.#2pt}}}
\def\square{\mathchoice\sqr64\sqr64\sqr{4.2}3\sqr{3.0}3}
\def\qed{\par\noindent\rightline{$\square$}}
\documentclass[a4paper,12pt]{article}
\usepackage[dvips]{graphicx}
\makeatletter
\renewcommand{\section}{{\setcounter{equation}{0}}\@startsection%
{section}%
{1}%
{0mm}%
{-\baselineskip}%
{0.5\baselineskip}%
{\normalfont\normalsize\bfseries}}%
\makeatother

\newcommand{\be}{\begin{equation}}
\newcommand{\ee}{\end{equation}}
\newcommand{\bea}{\begin{eqnarray}}
\newcommand{\eea}{\end{eqnarray}}
\newcommand{\non}{\nonumber}
\newtheorem{lem}{Lemma}[section]
\newtheorem{pro}{Proposition}[section]
\newtheorem{thm}{Theorem}[section]
\newtheorem{cor}{Corollary}[section]
\newtheorem{rmk}{Remark}[section]

\font\BB=msbm10
\def\RR{\hbox{\BB R}}
\def\NN{\hbox{\BB N}}

\def\ZZ{\hbox{\BB Z}}

\begin{document} 
\markboth{Edge states for a magnetic Hamiltonian}
{Edge states for a magnetic Hamiltonian}


%

\title{Propagating Edge States for a Magnetic Hamiltonian}

\author{Stephan De Bi\`evre\\ UFR de Math\'ematiques et UMR AGAT\\Universit\'e
des Sciences et Technologies de Lille\\ 59655 Villeneuve d'Ascq Cedex
France\\e-mail: debievre@gat.univ-lille1.fr 
\and 
Joseph V. Pul\'e\\ 
Department of Mathematical Physics\\
National University of Ireland, Dublin\\ 
(University College Dublin)\\
Belfield, Dublin 4, Ireland\\
e-mail: Joe.Pule@ucd.ie }

%
\date{February 23, 1999}


\maketitle

\begin{abstract}
We study the quantum mechanical motion of a charged particle moving in a 
half plane ($x>0$) subject to a uniform constant magnetic field $B$ directed
along the
$z$-axis and to an arbitrary impurity potential $W_B$, assumed to be weak in the
sense that $||W_B||_\infty < \delta  B$, for some $\delta$ small enough. 
We show rigorously a phenomenon pointed out by Halperin in 
his work on the quantum Hall effect, namely the existence of current
carrying and extended edge
states in such a situation. More precisely, we show that there exist
states propagating with a speed of size $B^{1/2}$ in
the $y$-direction, no matter how fast $W_B$ fluctuates. 
 As a result of this, we
obtain that the spectrum of the Hamiltonian is purely absolutely
continuous  in a spectral interval of size $\gamma B$ (for some $\gamma <1$)
between the Landau levels of the unperturbed system (i.e. the system without
edge or potential), so that the corresponding eigenstates are extended.  
\end{abstract}
\newpage
\section{Introduction}
It is well known that a classical charged particle, constrained to a plane and 
subjected to a perpendicular magnetic field will move along physical boundaries
when those are present. In the case of a particle moving in a half plane
($x>0, y\in\RR$), it is easy to see that the circular trajectories that are at
a distance less than $\sqrt E / B$ from the edge will bounce of it in such a way
that the particle speeds alongside the edge with a velocity of the order of
$\sqrt E$, where $E$ denotes the energy of the particle. If, on the other hand,
the centre of the trajectory is too far from the edge, it will not affect the
motion of the particle.

If, as one would expect, this picture is to carry over to the quantum
mechanical situation, then an initial state localized close to the edge in a
region of  size $B^{-1/2}$ -- an {\it edge state} -- should move ballistically
along the edge with a speed of order $\sqrt B$: here we used that the lowest
Landau level, in absence of the edge, is of order $B$. On the other hand,
although states further away from the edge -- {\it bulk states} --  should, due
to the uncertainty principle, not remain completely localized in the
$y$-direction, as in the classical case, they should nevertheless move much
more slowly than the edge states. This picture has long  been known to be
correct, but as a preparation for the case when an impurity potential is
present, we give a precise statement of the above properties in  Corollary
\ref{edgebulk}. We consider the Hamiltonian 
\begin{equation} 
H_0 =\frac{1}{2}p_x^2 +
\frac{1}{2}(p_y-Bx)^2,\label{1.a}
\end{equation}
with a Dirichlet boundary condition at $x=0$. 
Corresponding to each Landau band, we introduce the notion of 
$H_0$-invariant
{\it edge} and {\it bulk} spaces, with the following properties.
The
$y$-component of the velocity, given by $i[H_0,Y]$, is of order $\sqrt B$ on
an edge space, whereas it is exponentially small in $B$ on a bulk space
(Corollary \ref{edgebulk}). Furthermore, states belonging to the edge spaces
are negligeably small at distances much larger than the magnetic length
scale $1/\sqrt{B}$, reflecting the intuitively clear 
fact that the presence of the edge makes itself
felt only in a region of size $1/\sqrt B$ from the edge. In this sense the
edge states are quasi one-dimensional. The eigenfunctions of the restriction
of $H_0$ to the edge spaces are extended along the entire edge.

The existence of non-localized current-carrying quasi one-dimensional 
edge states plays a role in
certain theories of the quantum Hall effect \cite{h} (see \cite{gp}, 
\cite{w} and \cite{fs} for further details). It is therefore
of importance  to understand if such states exist in systems exhibiting a
quantized Hall resistance.  This is
argued to be the case  in  \cite{h}, in the case when the full Hamiltonian
is obtained by  adding a  weak impurity
potential to $H_0$. In other words, such potentials are not supposed 
to destroy the edge states existing in the free case.  
A very simple and rigorous proof of this statement is given in the 
present paper.  A
weak potential is a potential 
$W_B\in L^\infty(\RR_+\times\RR)$ satisfying 
$\delta_B \equiv ||W_B||_\infty < \frac{1}{2} B$. Since the distance between
successive Landau levels equals $B$, such a potential can not close the gaps 
between the Landau levels of the infinite system without a edge, even though its
size can be of order $B$: in this sense it is weak. It can however fluctuate
arbitrarily fast, and in particular on the magnetic length scale, which is of
order $1/\sqrt B$ (about $50-100$ Angstrom in realistic situations):
 this is important since, as explained in \cite{gp}, the weak
impurity potential is created by impurities at a distance of order $1/\sqrt B$
or less of the layer and can vary rapidly on this length scale.  As a typical
form for $W_B$ we can keep in mind a potential of the type
$$
W_B = \delta B \sum_{i\in\ZZ^+\times\ZZ} u_i(B^\alpha(\vec{x}- 
\frac{i}{B^\beta}))
$$
for some compactly supported site-potentials $u_i$ and exponents
$\alpha\geq 0$ and $\beta \geq 1/2$.

For weak potentials, we show that, in a spectral interval of size $B$
 between the Landau
levels, there are no bound states and that the speed in the $y$-direction is
still of order $\sqrt B$. 
As a consequence, we
obtain that in the same spectral interval, the spectrum is 
absolutely continuous, implying the corresponding eigenstates are extended.

The results described above in the case when no impurity potential is
present have been known for a long time and
can be obtained by studying explicitly the spectrum and eigenfunctions 
of
$H_0$, since it is an explicitly solvable Hamiltonian. Such an approach would
however not easily extend to the case when an impurity potential is added. 
Instead, we show in Proposition \ref{poscom2} that 
the magnitude of the speed in the 
$y$-direction
is strictly positive on the spectral subspaces corresponding to suitable 
spectral intervals between Landau levels. Such a positive commutator estimate is
then shown to be stable under perturbations in section 3, yielding the main
results via the virial theorem and the Mourre theory of positive commutators. 
(Theorem \ref{mainth}).

The idea that positive commutator methods and the virial theorem can be used to
obtain information about magnetic Hamiltonians in the presence of boundaries was
first proposed in \cite{mmp}. They consider a model with a {\it soft} edge, 
modeled
by a positive potential  $V$, supported on the negative axis and steeply rising
from $0$, and prove the absence of eigenvalues  in certain regions between the
Landau levels in this case. The conjugate operator used in this approach is the
quantum observable $C_y$ corresponding  to the $y$-coordinate of the centre of
the classical circular orbit: $C_y= y-(p_x/B)$. Classically this is indeed a
monotonic function of time for orbits close to the edge, since the Poisson
bracket $\{C_y, H\}=\frac{1}{B}(\partial_x V+\partial_x W)<0$ in that region,
provided the impurity potential $W$ has a small enough derivative.

In the present paper, we deal with the problem with a hard edge, as described
before.  We use the
$y$-coordinate itself as a conjugate operator, proving that 
the speed in the $y$-direction, $i[H,Y]$,  is strictly negative on edge states. 
This is marginally surprising since it is not true classically, but it
 turns out
to be extremely simple to understand in terms of the band structure
of the free Hamiltonian $H_0$.  
Using $y$ also has the important advantage of not introducing
derivatives of the potential in the commutator, as is the case when using
$C_y$, and therefore eliminating the need to control
their size. In addition, it renders the interpretation of the results in terms 
of propagation along the edge more transparent. On the down side, it is not 
obvious
the
present method will adapt itself easily to cases where the edge is not straight.

Let us  point out that we could treat the soft edge in the same way.
It seems however that this model does not lend itself to an analysis of the
high field regime, which is important for the quantum Hall effect. 
In that case, the particles will, even in the lowest Landau
level, penetrate deeply into the region $x<0$, so that there is an effective
edge around those values of $x$ where $V(x)\sim B$, where $V$ is the edge
potential. The high field behaviour of the speed, for example, will then depend
crucially on the precise shape of the edge, and this is not satisfactory.
We will therefore not deal any further with the soft edge in the following.

The results of \cite{mmp}  on the soft edge have recently been extended
\cite{fgw1} to a proof of
absence of singular continuous spectrum in suitable intervals
between the Landau levels, using the same conjugate operator as in \cite{mmp} 
to prove a positive commutator estimate. While 
finishing the present work, we learned that those results were further extended,
still using $C_y$ as a conjugate operator,
to the case of the {\it hard} edge in \cite{fgw2}. A result comparable to
our Theorem \ref{mainth} is proven there, but under the additional 
assumptions that both the first and second derivatives of the impurity
potential are small, so that rapid fluctuations in the impurity potential are
no longer allowed. In addition, our proof
is  technically
considerably less complicated partially because, in the model with a hard edge,
 the operator $C_y$ is symmetric but not self-adjoint, leading to complications
in applying the Mourre theory of positive commutators.


\section{The free Hamiltonian: edge and bulk spaces}
To study $H_0$ in (\ref{1.a}), we first use the translational invariance
in the $y$-direction to write
\begin{equation}
H_0=\int^\oplus_{\RR}dk \  H(k),\quad\hbox{with}\quad H(k) =
-\frac{1}{2}\frac{d^2}{dx^2} + \frac{1}{2}(k-Bx)^2, \ \ \ x>0,
\end{equation}
acting on $L^2(\RR_+\times \RR, dxdk)$, 
$k$ being the Fourier transform variable conjugate to $y$. 
We recall that $H_0$ is essentially self-adjoint on 
the space of functions $\varphi\in C_0^\infty (\overline{\RR_+} \times \RR)$ vanishing on the boundary \cite{he}.
\par

The spectrum of $H(k)$ consists of isolated non-degenerate 
eigenvalues $E_n(k)$, $n\in {\NN}_0$, with {\sl normalized} eigenfunctions 
$\varphi_n(x,k)$. We
will write ${\cal H}_n$ for the $n^{\hbox{th}}$ {\sl band space}, namely the 
space consisting of
vectors of the form $f(k)\varphi_n(x,k), f\in L^2(\RR, dk)$. This is an
$H_0$-invariant subspace of $L^2(\RR_+\times \RR, dxdk)$; we shall on
occasion view it as a subspace of  $L^2(\RR_+\times\RR, dxdy)$ as well, with the 
same notation. To understand the behaviour of
the $E_n(k)$ and the $\varphi_n(x,k)$, and in particular their dependence on
$B$, we introduce the following scaling:
\begin{equation}
\tilde x = \sqrt B x, \ \tilde y= \sqrt B y, \ H_0=B{\tilde H_0},\  
{\tilde H_0}=-\frac{1}{2}\frac{\partial^2}{\partial\tilde x^2} +
\frac{1}{2}(\frac{1}{i}\frac{\partial}{\partial\tilde y}-\tilde x)^2.
\end{equation}
Note that, strictly speaking, $H_0$ is unitarily equivalent to $B{\tilde H_0}$, 
not 
equal to it, but since the unitary transformation is just the rescaling
of the variables, we allow ourselves this slight abuse of notation. Again
\begin{equation}
{\tilde H_0}=\int^\oplus_{\RR} d\kappa\ {\tilde H}(\kappa),
\quad \hbox{with}\quad {\tilde H}(\kappa) =
-\frac{1}{2}\frac{d^2}{d\tilde x^2} 
+ \frac{1}{2}(\kappa-\tilde x)^2, \ \ \ \ \tilde x>0.
\end{equation}

Here $\kappa$ is the Fourier transform variable conjugate to $\tilde y$, so
that $ky=\kappa\tilde y$ and hence $k=\sqrt B \kappa$. The spectrum of
${\tilde H}(\kappa)$ consists of isolated eigenvalues $\alpha_n(\kappa)$. The 
normalized eigenfunctions
of ${\tilde H}(\kappa)$, at each fixed $\kappa$, $\tilde \varphi_n (\cdot, 
\kappa)$ are given by
\bea
\tilde \varphi_n (\tilde x, \kappa)=C_n 
D_{\alpha_n(\kappa)-1/2}(\sqrt2({\tilde x}-\kappa)),
\eea
where $D_{\alpha-1/2}$ is the Whittaker function (\cite{as} p686) with parameter 
$\alpha$ and $\alpha_n(\kappa)$ is determined by the boundary condition 
\be
D_{\alpha_n(\kappa)-1/2}(-\sqrt2\kappa)=0.
\label{2.b}
\ee
One can check that the eigenvalues $\alpha_n(\kappa)$ 
are smooth functions of $\kappa$. The other properties of the
$\alpha_n(\kappa)$ that we shall be needing are collected in the following 
Lemma:

\begin{lem} \label{alphan} 
\begin{itemize}
\item[(i)] $\alpha_n(0)=2n +3/2,
\qquad{\displaystyle \alpha'_n(0)=
-\frac{(2n+2)!}{n!(n+1)!\pi^{1/2}2^{2n}}},$
\item[(ii)]$\alpha'_n(\kappa)=
-\frac{1}{2}|\tilde \varphi_n'(0,\kappa)|^2<0,$
\item[(iii)] $\alpha_{n}(\kappa)>(n+\half)$ and there exist $C_n>0$ 
so that, for all $\kappa\geq 0$, 
$$
\alpha_{n}(\kappa) -(n+\frac{1}{2}) \leq 
C_n \exp -\frac{1}{4}(\kappa -\sqrt n)^2.
$$
\item[(iv)]  For all $n\in \NN_0$ and all $\epsilon>0$, there exist 
positive constants $X_n$, $K_{n,\epsilon}, C_{n,\epsilon}$ so that, for all
$\kappa\geq K_{n,\epsilon}$ and all $\tilde x\in[0,X_n]$,
$$
|\tilde\varphi_n(\tilde x, \kappa)|^2\leq
C_{n,\epsilon}\exp\left \{-\half (1-\epsilon)(\tilde x - \kappa)^2\right \},
$$
and
$$ 
|\alpha_n'(\kappa)| \leq C_{n,\epsilon}\exp \left \{-\half 
(1-\epsilon)\kappa^2\right \}.
$$
\item[(v)] For all $\kappa <0$, $|\alpha_n'(\kappa)|> |\kappa|$ 
and
$\lim_{\kappa\to-\infty}\alpha_n(\kappa) =\infty$.
\end{itemize}
\end{lem}

The proof of this lemma, which uses only  standard techniques of Schr\"odinger 
operator theory, is postponed to Section \ref{proof}.  Numerical computation of
the $\alpha_n(\kappa)$ indicates that they are convex functions of 
$\kappa$ (see Figure 1). In addition, it seems that
$$
\alpha_{n+1}(\kappa) - \alpha_n(\kappa) >1,
$$
a result that is  intuitively clear, at least for $\kappa>0$. 
We have not been able to prove these last two results and shall not use them.  
If
they are true, the statements of our results below can be simplified somewhat. 

%

\begin{figure}[hbt] 

\begin{center}\includegraphics[width=10cm]{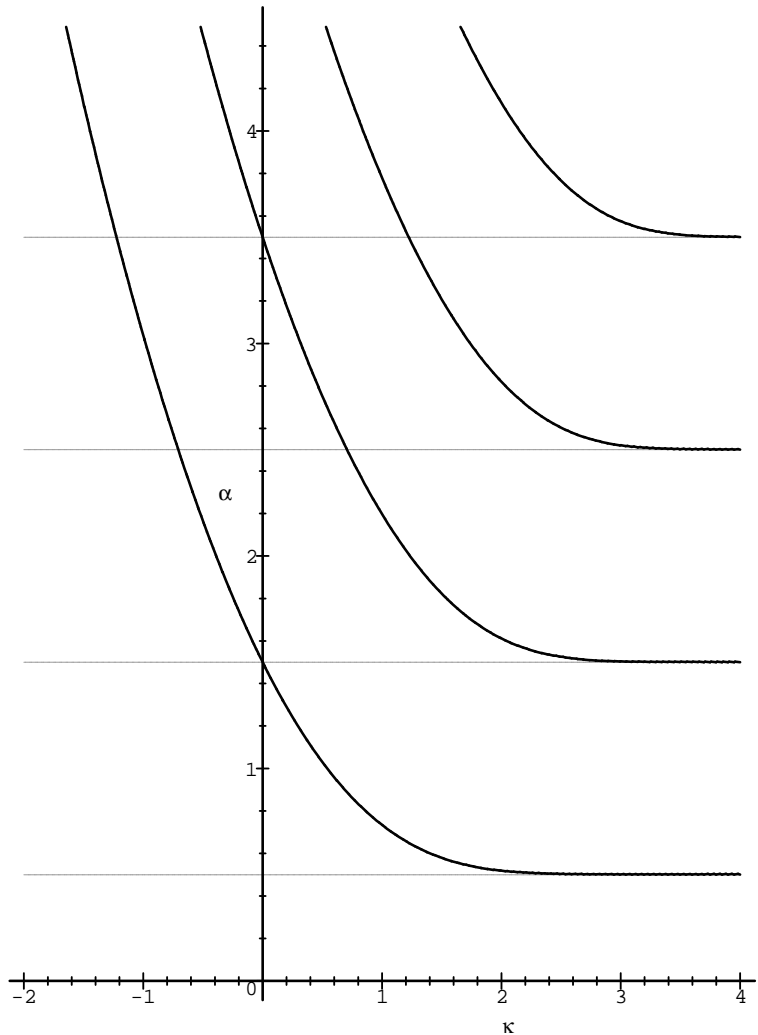}


\end{center}


\caption{$\alpha_n(\kappa)$ for $n=0,1,2,3$}

\label{ }

\end{figure}

\par
It is clear from the above that the spectrum of ${\tilde H}_0$ and hence of
$H_0$ is absolutely continuous and fills the entire half-axis from $1/2$ to
infinity. The bands $E_n(k)$ can now be written
\begin{equation}
E_n(k) = B\alpha_n(\frac{k}{\sqrt B}).
\end{equation}

Writing $\varphi_n (x, k)$ for the normalized eigenfunctions
of ${H}(k)$ (at each fixed $k$), we have 
\begin{equation}
\varphi_n(x,k) = B^{1/4} \tilde\varphi_n(\sqrt B x, \frac{k}{\sqrt B}).
\label{eigenf}\end{equation}
These simple observations will now allow us to define within
${\cal H}_n$   {\it  edge spaces} and  {\it bulk spaces} as follows. We define,
for each
$\sigma>0,\gamma>0$:
\begin{eqnarray}
{\cal H}_{n,e}(\sigma, \gamma) &\cong& L^2(]-\infty, \sigma B^\gamma], dk)
\subset {\cal H}_n,\\
{\cal H}_{n,b}(\sigma, \gamma) &\cong& L^2([ \sigma B^\gamma, \infty), dk)
\subset {\cal H}_n,\\
{\cal H}_n& =& {\cal H}_{n,e} \oplus {\cal H}_{n,b}.
\end{eqnarray}
Note that these spaces are $H_0$ invariant. We will call ${\cal H}_{n,e}(\sigma,
\gamma)$ an edge space for all
$\gamma\leq 1/2$ and ${\cal H}_{n,b}(\sigma, \gamma)$ a bulk space for all
$\gamma>1/2$. 
For a different approach to the definition of bulk and edge spaces,
in the case of a bounded geometry, we refer to \cite{av}.

To understand those definitions, recall first that
a standard  stationary phase argument shows
that $-\partial_k E_n(k_0)=-\sqrt B\alpha_n'(k/{\sqrt B})$
 is the  group speed in the $y$-direction
of a wave packet $f(k)\varphi_n(x, k)$ with the support of $f$ close to $k_0$.
If $k_0$ is inside an interval $(-\infty, k_B]$ where $k_B$ is of order $\sqrt B$ 
or smaller,  the wave packet belongs to 
the edge space ${\cal H}_{n,e}(\sigma, 1/2)$ and it follows from Lemma
\ref{alphan}  that such a wave packet speeds along the edge in the $y$ direction
with a velocity of order $\sqrt B$. In addition, it follows from standard
exponential estimates on the eigenfunctions $\tilde\varphi_n$ (as in the proof of Lemma \ref{alphan}) that in this
case the wave packet is exponentially small for $x$ much bigger than $1/\sqrt
B$.  If, on the other hand, $k_0$ belongs to an interval of the form $[k_B,\infty[$
with $k_B$ of order $B^\gamma, \gamma >\half$, then the group velocity is 
exponentially small in $B$ (see Lemma \ref{alphan} (iv)). In addition, if 
$f(k)\varphi_n(x,k)\in {\cal H}_{n,b}(\sigma, \gamma)$, with $\gamma>\half$, then 
Lemma \ref{alphan}(iv) immediately implies that 
$$
\int_0^{\frac{1}{\sqrt B}} \int_{-\infty}^{\infty}|f(k)\varphi_n(x,k)|^2
dkdx \leq C_{n,\epsilon} \exp -(1-\epsilon)(\sigma^2 B^{2\gamma -1}-1),
$$
so that the wave packet is exponentially small in the region 
$0\leq x\leq \frac{1}{\sqrt B}$ close to the edge.
We note also that  the spectrum of
$H_0$ restricted to a bulk space ${\cal H}_{n,b}(\sigma, \gamma)$ is an
exponentially small interval (in $B$) just above the $n$th Landau level (Lemma \ref{alphan} (iii)),
that we will refer to as the bulk spectrum. The spectrum of $H_0$ restricted to
an edge space ${\cal H}_{n,e}(\sigma, 1/2)$ -- {\em the edge spectrum} -- is on
the other hand of the form 
$[B(n+\frac{1}{2} + c_\sigma),\infty)$. In particular, it fills up an interval 
of
size
$B$ below the $(n+1)$th Landau level, including the latter. 

To give a formulation of the above statements that is at once more precise and
does not use the band structure of the Hamiltonian $H_0$, so that it has a
chance to pass to the perturbed Hamiltonian, we now turn  to the statement
and proof of
 a {\it positive commutator estimate}. We will show that the speed $V_y =
i[H_0, Y]$, where the operator $Y$ is multiplication by $y$,
is strictly negative away from the Landau levels. This is the content of the 
following proposition, which will be
generalized to the perturbed Hamiltonian in the next section.  
\par
Let $L_n=(n+1/2,n+3/2]$ be the $n$th Landau band when $B=1$.
Let
\be
\theta_n(\kappa,n',n'')
=\cases{| \alpha_{n'}(\kappa)- \alpha_{n''}(\kappa)|& if both 
$\alpha_{n'}(\kappa)$ and $\alpha_{n''}(\kappa)$ are in $L_n$,\cr
1 & otherwise. \cr}
\ee
We let $\delta_0=1$ and for 
$n \geq 1$ we let
\be
\delta_n=\inf_{{n'\neq n''}\atop{n',n''\leq n}}
\inf_\kappa \theta_n(\kappa,n',n'').
\label{2.c}
\ee
Note that for $n'>n$, $\alpha_{n'}(\kappa)>n+3/2$ for all $\kappa$, 
so that in (\ref{2.c}) it is not necessary to have the restriction
$n',n''\leq n$. From Lemma \ref{alphan} {\it(i)} we see that $\alpha_0(0)=3/2$ 
and $\alpha_1(0)=7/2$ and from {\it(ii)} we see that $\alpha_n(\kappa)$ is 
strictly decreasing in $\kappa$.
Therefore for $\kappa \leq 0$ only $\alpha_0(\kappa)$ can be in $L_1$ while for
$\kappa > 0$ only $\alpha_1(\kappa)$ can be in $L_1$.
Definition (\ref{2.c}) then  implies that
$\delta_1=1$ also.  As mentioned before, Fig.1 suggests that  $\delta_n=1$ for
all $n$. We will only prove $\delta_n>0$, as follows.

\par

Since $\alpha_{n'}(\kappa)\to \infty$ as $\kappa \to -\infty$ for all
$n'$, there exists $\kappa_n<0$ such that  for all $\kappa<\kappa_n$,
$\alpha_{n'}(\kappa)>n+3/2$ for $n'\leq n$. 
Also there exists $\kappa'_n>0$ such that
for all $\kappa>\kappa'_n$, $\alpha_{n'}(\kappa)<n+1/2$ for $n'< n$. 
Thus for $n'\neq n''$, $n',n''\leq n$, $\theta_n(\kappa,n',n'')=1$ for $\kappa$ 
outside the compact interval 
$[\kappa_n,\kappa'_n]$ and therefore $\delta_n>0$.
\par\noindent
\begin{pro}\label{poscom2}
Let $\Delta \subset L_n $ be a closed interval with $|\Delta|<\delta_n$. Let 
\be
\nu_-(\Delta)= \inf_{\{n',\kappa \ |\ \alpha_{n'}(\kappa)\in \Delta \}}| 
\alpha'_{n'}(\kappa)| 
>0
\ee
and
\be
\nu_+(\Delta)= \sup_{\{n',\kappa \ |\ \alpha_{n'}(\kappa)\in \Delta \}}| 
\alpha'_{n'}(\kappa)| 
>0.
\ee
If ${\tilde Y}$ is multiplication by ${\tilde y}$ and ${\tilde P}_0(\Delta)$
is the spectral projection of ${\tilde H}_0$ onto $\Delta$, then
\begin{equation}
\nu_-(\Delta){\tilde P}_0(\Delta)\leq
{\tilde P}_0(\Delta) i[{\tilde Y}, {\tilde H}_0]{\tilde P}_0(\Delta)\leq
\nu_+(\Delta){\tilde P}_0(\Delta),
\end{equation}
and consequently, for $\psi$ with $||\psi||=1$ in the range of ${\tilde 
P}_0(\Delta)$
\begin{equation}
-\nu_+(\Delta)t \leq \langle\psi_t, {\tilde Y} \psi_t\rangle - \langle\psi_0, 
{\tilde Y}
\psi_0\rangle \leq -\nu_-(\Delta)t.
\end{equation} 
\end{pro}
\noindent{\bf Proof:} We have that 
$\alpha_{n'}^{-1}(\Delta)\cap\alpha_{n''}^{-1}(\Delta) =\emptyset $ 
if $n'\neq n''$, since $|\Delta|< \delta_n$.
Thus for any $\psi$ we can write
\be
{\tilde P}_0(\Delta)\psi(\tilde x, \kappa)
= \sum_{n'=0}^n\psi_{n'}(\tilde x, \kappa),
\ee
where 
\be
\psi_{n'}({\tilde x}, \kappa)=\beta_{n'}(\kappa)
{\bf 1}_{\alpha_{n'}^{-1}(\Delta)}(\kappa)
{\tilde\varphi}_{n'}({\tilde x}, \kappa)
\ee
and
\be
\beta_{n'}(\kappa) =\int_{\RR_+} \overline{\psi(\tilde x,\kappa)}\ \tilde
\varphi_{n'}(\tilde x,\kappa) dx .
\ee
Since $i[{\tilde Y}, {\tilde H}_0] = {\tilde x}-p_{\tilde y}$ , 
it is clear that $\langle \psi_{n'}, i[{\tilde Y}, {\tilde H}_0]
\psi_{n''}\rangle=0$ if $n'\not=n''$ since the supports of $\psi_{n'}$ and of 
$\psi_{n''}$ 
are disjoint in the $\kappa$ variable. On the other hand,
\be
\langle \psi_{n'}, i[{\tilde Y}, {\tilde H}_0]\psi_{n'}\rangle
=\int_{\alpha_{n'}^{-1}(\Delta)}d\kappa\
|\beta_{n'}(\kappa)|^2\int_{\RR_+}d{\tilde x}  ({\tilde
x}-\kappa)|\varphi_{n'}({\tilde x},\kappa)|^2 ,
\ee
and, by the Feynman-Hellman theorem
$$
\int_{\RR_+}d{\tilde x} ({\tilde x}-\kappa)|{\tilde\varphi}_{n'}({\tilde 
x},\kappa)|^2
= -\alpha'_{n'}(\kappa) = |\alpha'_{n'}(\kappa)|.
$$
The 
proposition is now immediate. 
\qed
\par
Using the scaling behaviour in $B$ we now have the following Corollary:
\begin{cor}\label{edgebulk}\mbox{}\hfil\break 
(i) Let $n\in\NN$ be fixed and let 
$\Delta\subset ((n+\frac{1}{2})B, (n+\frac{3}{2})B]$ be a closed interval
with $|\Delta|<\delta_n B$. Then
\be
\sqrt B\nu_-(B^{-1}\Delta)P_0(\Delta)\leq
P_0(\Delta) i[Y, H_0]P_0(\Delta)\leq
\sqrt B\nu_+(B^{-1}\Delta)P_0(\Delta).
\ee
where $B^{-1}\Delta=\{E/B \ |\ E\in \Delta\}$. 
\par
\noindent (ii) For all $n\in\NN$, for all $\sigma>0$ there exists a 
constant $C_{n, \sigma}>0$ so that for all 
$\psi \in {\cal H}_{n,e}(\sigma,1/2)$ and for all $B$
\begin{equation}
\langle\psi, i[Y,  H_0]\psi\rangle 
\geq\sqrt B \inf_{\kappa\leq \sigma }|\alpha'_n(\kappa)|\ ||\psi ||^2
 \ >C_{n,\sigma}\sqrt B ||\psi ||^2.
\end{equation}
(iii)  Let $\epsilon>0$. Then for all $n\in{\NN}_0$, 
for all $\sigma>0$ there exists a constant $C_{n,\sigma,\epsilon}>0$ so that for
all $B$ and for all $\psi \in {\cal H}_{n,b}(\sigma,1/2+\epsilon)$
\begin{equation}
\langle\psi, i[Y, H_0]\psi\rangle\leq
\sqrt B \sup_{\kappa\geq \sigma B^\epsilon }|\alpha'_n(\kappa)|\ ||\psi ||^2
\ < C_{n,\sigma,\epsilon}\sqrt B\exp\left
\{-\half(1-\epsilon)\sigma^2B^{2\epsilon}\right \}||\psi||^2.
\end{equation}
\end{cor}
\noindent{\bf Proof:} This is now an immediate consequence of Lemma
\ref{alphan} and of the proof of Proposition \ref{poscom2}.\qed

\begin{rmk} Parts (ii) and (iii) of the corollary state that the speed
in the y direction is at least of order $\sqrt B$  for any edge state and
at most of order  $\exp-B^{2\epsilon}$ for any bulk state.  
\end{rmk}


\section{Adding a weak impurity potential}
We now consider the Hamiltonian 
$$
H=H_0 + W_B
$$
where $W_B\in L^\infty(\RR_+\times \RR, dxdy)$ is a real potential satisfying 
$||W_B||_\infty \leq AB$ where $A<\infty$ is independent of $B$.  
Let
$$
{\tilde H}={\tilde H}_0 + \tilde W_B ; \tilde W_B(\tilde x, \tilde y)
=B^{-1}W_B(\frac{\tilde x}{\sqrt B}, \frac{\tilde y}{\sqrt B})
$$
and let ${\tilde P}(\cdot)$ denote the spectral family of ${\tilde H}$.
\par
Our main result is then the
following theorem, which should be compared to Proposition 2.1.
For $\lambda <1$, let $L_n^\lambda = (n+1/2+\lambda, n+3/2]$.
Let 
\be
\nu(n,\lambda)=\nu_-(L_n^\lambda)
= \inf\{| \alpha'_{n'}(\kappa)| \ |\  n'\leq n,\
n+1/2+\lambda<\alpha_{n'}(\kappa)\leq n+3/2 \}  >0.
\label{nunlambda}\ee
\begin{thm}\label{mainth} Let
$n\in\NN$ be fixed. Let
$\lambda$,
$\lambda' >0$ with
$\lambda +\lambda'<1$ and let 
$L_n^{\lambda,\lambda'}= (n+1/2+\lambda, n+3/2-\lambda')$.
There exists $\delta(n,\lambda,\lambda')>0$ such that if 
$||W_B||_\infty<\delta(n,\lambda,\lambda') B$
and
$\epsilon<\delta(n,\lambda,\lambda')$,  then, for all $\alpha \in L_n^{\lambda,\lambda'}$, for the
interval 
$\Delta \equiv (\alpha-\epsilon, \alpha+\epsilon)$,
\be
{\tilde P}(\Delta) i[{\tilde Y}, {\tilde H}]{\tilde P}(\Delta)\geq
\half \nu(n, \lambda/2){\tilde P}(\Delta).
\label{3.d}
\ee
Consequently if $||W_B||_\infty<\delta(n,\lambda, \lambda')B$, then
\be
\sigma_{\rm sing}({\tilde H})\cap L_n^{\lambda,\lambda'} =\emptyset.
\label{3.e}
\ee
\end{thm}
\par 
Clearly we can give a scaled up version of this theorem:
\begin{cor}\label{maincor}Let $n\in\NN$ be fixed and let $\lambda$, $\lambda' 
>0$ with
$\lambda +\lambda'<1$. There exists $\delta(n,\lambda,\lambda')>0$ 
such that if $||W_B||_\infty<\delta(n,\lambda,\lambda') B$, then
\be
\sigma_{\rm sing}(H)\cap (B(n+1/2+\lambda), B(n+3/2-\lambda')) =\emptyset.
\ee
\end{cor}
It is useful to have the following variant of Theorem \ref{mainth}. Here 
we fix a bound on $||W_B||_\infty/B$ and give the dependence on this bound 
of the endpoints $a$,$b$ of the interval $(a,b)\subset L_n$, such that 
$(a,b)$ contains only absolutely continuous spectrum.
\begin{thm}\label{varth} Let
$n\in\NN$ be fixed. 
Suppose that 
$||W_B||_\infty<\delta B$ where $\delta<1/2$.
Let
$\lambda_n$,
$\lambda_n' \in (0,1/2)$ be such that
$\min(\lambda_n,\delta_n) \nu(n,\lambda_n/2)^2 > 2^9(n+2)\delta$ and 
\break
$\min(\lambda_n',\delta_n) \nu(n,1/4)^2 > 2^9(n+2)\delta$
then 
for all $\alpha \in L_n^{\lambda_n,\lambda_n'}$, there exists an interval 
$\Delta $ containing $\alpha$ such that
\be
{\tilde P}(\Delta) i[{\tilde Y}, {\tilde H}]{\tilde P}(\Delta)\geq
\half \nu(n, \lambda_n/2){\tilde P}(\Delta).
\label{3.f}
\ee
Therefore
\be
\sigma_{\rm sing}({\tilde H})\cap L_n^{\lambda_n,\lambda_n'} =\emptyset.
\label{3.g}
\ee
\end{thm}
\par 
Note that, given $\delta$, no $\lambda_n$ and $\lambda_n'$ satisfying the 
conditions of the theorem might exist. Nevertheless, it is clear that for 
sufficiently small $\delta$, the above results guarantee the existence of an 
interval of absolutely continuous spectrum between the Landau levels.
The scaled up version of this theorem is then:
\begin{cor}\label{varcor} Under the conditions of Theorem \ref{varth},
\be
\sigma_{\rm sing}(H)\cap (B(n+1/2+\lambda_n), B(n+3/2-\lambda_n')) 
=\emptyset.
\ee
\end{cor}
\par\noindent
{\bf Proof of Theorem \ref{mainth}:}\ Note first that 
$i[\tilde Y, \tilde H_0] = i [\tilde Y, \tilde H]$,
so that the result would follow from Proposition 2.1 if we could
replace ${\tilde P}(\Delta)$ by ${\tilde P}_0(\Delta)$. This
can indeed be achieved with a few tricks and at not too high 
a cost, provided
one replaces the interval $\Delta$ by an auxiliary one $\Delta'$,
that is larger but for which $\nu(\Delta')$ is not too small. 
Let $\sigma\equiv \min(\lambda, \lambda', \delta_n)/4$, where 
$\delta_n$ is
defined in (\ref{2.c}), and let
$\Delta'$  be the interval $[\alpha-\sigma, \alpha+\sigma]\subset
L^{\lambda/2,\lambda'/2}_n$.
Let $\Delta$
be the interval $[\alpha-\epsilon,\alpha+\epsilon]$,
where $\epsilon\leq\sigma$. 
Let $\psi\in {\tilde P}(\Delta){\cal H}$.
Then, recalling that $A \leq ||\tilde W_B||_\infty$,
$$
||({\tilde H}_0 - \alpha)\psi||\leq ||({\tilde H}-\alpha){\tilde P}
(\Delta)\psi||
+A||\psi||\leq(\epsilon + A)||\psi||.
$$
Hence
\be
||{\tilde P}_0(\Delta^{'c})\psi||\leq
||\frac{1}{{\tilde H}_0-\alpha}\tilde P_0(\Delta^{'c})||
\ ||({\tilde H}_0-\alpha)\psi||\leq
{\sigma}^{-1}(\epsilon + A)||\psi||,
\label{3.c}
\ee
since
$\hbox{min}\{|\lambda - \alpha|\ |\ \lambda\in\Delta'^{c}\}\geq \sigma.$ 
Clearly
\begin{eqnarray}
i\langle\psi, [\tilde Y, \tilde H]\psi\rangle
&\geq &i\langle \tilde P_0(\Delta')\psi,[\tilde Y, \tilde H_0]
\tilde P_0(\Delta')\psi\rangle\nonumber\\
&\ & \hskip 1cm -2||[\tilde Y, \tilde H_0]\tilde P_0 (\Delta^{'c})\psi||\ ||\psi||.
\label{3.a}
\end{eqnarray}
The required positivity will come from the first term, 
so we only have to control the last one. We find
\begin{eqnarray*}
||[{\tilde Y}, {\tilde H}_0]{\tilde P}_0(\Delta^{'c})\psi|| 
&\leq&2 \langle
{\tilde P}_0(\Delta^{'c})\psi, {\tilde H}_0 {\tilde P}_0
(\Delta^{'c})\psi\rangle^{1/2}\\
&\leq&
2||{\tilde H}_0{\tilde P}_0(\Delta^{'c})\psi||^{1/2} 
||{\tilde P}_0(\Delta^{'c})\psi||^{1/2}.
\end{eqnarray*}
But
\begin{eqnarray*}
||{\tilde H}_0{\tilde P}_0(\Delta^{'c})\psi||^{1/2}
&\leq& (||{\tilde H}\psi|| + A||\psi||)^{1/2}\\
&\leq&( n+3/2+A)^{1/2}||\psi||^{1/2}\\
&\leq&( n+2)^{1/2}||\psi||^{1/2},
\end{eqnarray*}
if $A\leq 1/2$. Therefore
$$
||[{\tilde Y}, {\tilde H}_0]{\tilde P}_0(\Delta^{'c})\psi||\leq 
2( n+2)^{1/2}{\sigma}^{-1/2}(\epsilon +A)^{1/2}||\psi||.
$$
Inserting this into (\ref{3.a}) yields 
$$
\hskip-2cm i\langle \psi, 
[{\tilde Y}, {\tilde H}]\psi\rangle 
\geq i\langle {\tilde P}_0(\Delta')\psi,
[{\tilde Y}, {\tilde H}_0] 
{\tilde P}_0(\Delta')\psi\rangle
$$
\be
 \hskip2cm - 4( n+2)^{1/2}{\sigma}^{-1/2}(\epsilon +A)^{1/2}||\psi||^2.
 \label{3.j}\ee
On the other hand, since $|\Delta'|=2\sigma <\delta_n$, 
Proposition 2.1 states that
$$
i\langle {\tilde P}_0(\Delta')\psi, [{\tilde Y}, 
{\tilde H}_0]{\tilde P}_0(\Delta')\psi\rangle \geq
\nu(\Delta')||{\tilde P}_0(\Delta')\psi||^2
\geq
\nu(n,\lambda/2)||{\tilde P}_0(\Delta')\psi||^2,
$$
where $\nu(n,\lambda/2)$ is defined in (\ref{nunlambda}). 
Inserting this into (\ref{3.j}) and using (\ref{3.c}) together with the observation that
$
||\psi||^2=||{\tilde P}_0(\Delta')\psi||^2+||{\tilde P}_0(\Delta'^{c})\psi||^2,
$
 yields
\be
 i\langle \psi,[{\tilde Y}, {\tilde H}]\psi\rangle 
\geq \nu(n,\frac{\lambda}{2})
\bigl[1-\left(\frac{(\epsilon+A)^2}{\sigma^2}+
\frac{4(n+2)^{1/2}(\epsilon +A)^{1/2}}{{\sigma}^{1/2}
\nu(n,\frac{\lambda}{2})}\right)\bigr]||\psi||^2.
\label{ineq}\ee
Let $\delta(n,\lambda,
\lambda') =\min\left (\frac{\sigma \nu(n,\lambda/2)^2}{
2^9(n+2)}, 
\frac {\sigma}{4}, \frac{1}{2}\right)$. 
Then if $A<\delta$ and $\epsilon <\delta$, one has that 
$4( n+2)^{1/2}{\sigma}^{-1/2}(\epsilon +A)^{1/2}
<\frac{1}{4}\nu(n,\lambda/2)$ and $((\epsilon + A)/\sigma)^2\leq 1/4$, so that
 the first statement in the theorem
follows. To prove  (\ref{3.e}) 
it is now sufficient to use (\ref{3.d}) and to apply the Mourre theory of
positive commutators. For a textbook treatment, we refer to  \cite{abg}; see
also \cite{gg} for a concise review of the domain questions involved. The latter 
are trivial in the present case. Indeed, the commutator $[H_0, Y] = [H, Y]$ is 
obviously relatively $H_0$ bounded, the domain of the Hamiltonian is invariant 
under the unitary group $\exp is Y$ and the second commutator $[[H_0,Y],Y]$ is
bounded.
\qed


\section{Proof of Lemma \ref{alphan} }\label{proof}
\par
Part (i) follows from a computation using standard
properties of the Hermite polynomials.  To prove (ii), we write $V_\kappa(\tilde
x) = \half (\tilde x - \kappa)^2$ and use the Feynman-Hellman formula  to write
(see \cite{hd})
\begin{eqnarray*}
\alpha'_n(\kappa)&=& -\int_0^\infty V_\kappa'(\tilde x) 
\tilde\varphi_n^2(\tilde x, \kappa) \ d\tilde x\\
&=&2\int_0^\infty V_\kappa (\tilde x) \tilde \varphi_n (\tilde x, \kappa) 
\tilde \varphi_n'(\tilde x, \kappa) \ d\tilde x\\
&=& \int_0^\infty \varphi_n'' (\tilde x, \kappa)\varphi_n' (\tilde x, \kappa) 
\ d\tilde x + 2\alpha_n(\kappa) 
\int_0^\infty \varphi_n(\tilde x, \kappa)\varphi_n' (\tilde x, \kappa) \ d\tilde x,
\end{eqnarray*}
from which the result follows.  Note that by uniqueness $\varphi_n'(0)$ cannot be zero. 
For (iii), we will use a perturbative argument, treating
the Dirichlet boundary condition at $0$ as a perturbation. We note first that, 
by
the min-max principle,  $\alpha_n(\kappa)> n+\half$. Now, let $h_n$ denote the
$n$th  Hermite function and let $h_{n,\kappa}(x)=h_n(x-\kappa)$. 
Let $\theta$ be a smooth function such that $\theta(x)=0$ for $x\leq 0$ and
$\theta(x)=1$ for $x\geq 1$ We compute
$$
(\tilde H (\kappa) - (n+\half))\theta h_{n,\kappa}=[\tilde H(\kappa), 
\theta]h_{n,\kappa}=
\half(-\theta''-2i\theta'p)h_{n,\kappa}.
$$
Now, since the supports of $\theta'$ and $\theta''$ are contained
in $[0,1]$, and since 
$$
||\theta'p h_n||^2 = \langle h_{n,\kappa}, [p,\theta'^2] p h_{n,\kappa}\rangle +
\langle \theta'^2h_{n,\kappa}, p^2h_{n,\kappa}\rangle,
$$
one easily concludes there exists a constant $C_n$ so that
$$
||(\tilde H(\kappa) - (n+\half))\theta h_{n,\kappa}||
\leq C_n 
||{\bf 1}_{[0,1]} h_{n,\kappa}||^{\half},
$$
where ${\bf 1}_{[0,1]}$ denotes the characteristic function
of $[0,1]$. Standard properties of the Hermite functions then imply that, for 
$\kappa$ large enough
$$
||(\tilde H(\kappa) - (n+\half))\theta h_{n,\kappa}||
\leq C_n
\exp-\frac{1}{4}(\kappa-\sqrt n)^2.
$$
This shows that, for $n$ fixed, 
$$
{\rm dist}(\sigma(\tilde H(\kappa)),(n+\half))
\leq 2C_n
\exp-\frac{1}{4}(\kappa-\sqrt n)^2.
$$
For $n=0$, $|\alpha_0(\kappa)-\half|={\rm dist}(\sigma(\tilde H(\kappa)),
\half)\leq 2C_0\exp-\frac{1}{4}(\kappa)^2$, since $\alpha_1(\kappa)>3/2$, and 
(iii) 
then follows by induction on $n$.
We now turn to the proof of (iv). This only involves a rather straightforward
application of the standard method for proving exponential decay estimates on
eigenfunctions in a classically  forbidden region (see, for example \cite{a,
he}). With $V_\kappa(\tilde x)\equiv\half (\tilde x-\kappa)^2$ as before,
we first define, for all $\kappa>0$, $0<x_n(\kappa)<\kappa$ by
$V_\kappa(x_n(\kappa))=\alpha_n(\kappa)$. 
Clearly, for all
$0\leq
\tilde x\leq x_n(\kappa)$, $\tilde\varphi_n(\tilde x, \kappa)$ and 
$\tilde\varphi_n''(\tilde x, \kappa)$ have the same sign, which we can 
assume to be strictly positive. 
Also, on the same region $\varphi_n'(\tilde x,
\kappa)>0$.  As a result, for any $a\in[0,x_n(\kappa)-2]$, one has
$$
|\tilde\varphi_n(a,\kappa)|^2\leq \int_a^{a+1}
|\tilde\varphi_n(y,\kappa)|^2 dy.
$$
Let, for $0\leq \tilde x\leq x_n(\kappa)$,
$$
f_n(\tilde x,\kappa) = \int_{\tilde x}^{x_n(\kappa)}
\sqrt{2(V_\kappa(y)-\alpha_n(\kappa)+1)} dy.
$$
Note that
$$
\half f'_n(\tilde x,\kappa)^2 - 
(V_\kappa(\tilde x)-\alpha_n(\kappa)+1)=0.
$$
We introduce $\eta_n(\tilde x, \kappa)$, a smooth characteristic function of the
interval $[0, x_n(\kappa)-2]$, with ${\rm supp}\ \eta_n'\subset[x_n(\kappa)-2,
x_n(\kappa)-1]$.  Then
\bea
\int_a^{a+1} |\tilde\varphi_n(y,\kappa)|^2 dy &\leq&
\exp-2f_n(a,\kappa)\int_a^{a+1}\exp 2f_n(y,\kappa) |\tilde\varphi_n
(y, \kappa)|^2 dy\non \\
&\leq &\exp-2f_n(a,\kappa)
\langle\psi_n,(V_\kappa-\alpha_n(\kappa)-\half {f'}_n^2) \psi_n\rangle,
\eea
where $\psi_n = \eta_n (\exp f_n) \tilde\varphi_n$.  
A simple computation shows
$$
(\exp f_n)(H-\alpha_n)(\exp-f_n) \psi_n= \frac{p^2}{2}\psi_n + (V_\kappa
-\alpha_n-\half {f'}_n^2)\psi_n + \half(f'_n\frac{d}{d\tilde x} + 
\frac{d}{d\tilde x}f'_n)\psi_n,
$$
so that
$$
\hbox{\rm Re}\langle\psi_n, (\exp f_n)(H-\alpha_n)(\exp-f_n)\psi_n\rangle
\geq\langle \psi_n, (V_\kappa-\alpha_n-\half {f'}_n^2)\psi_n\rangle.
$$
On the other hand
$$
\hbox{\rm Re}\langle\psi_n, (\exp f_n)(H-\alpha_n)(\exp-f_n)\psi_n\rangle =
\hbox{\rm Re}\langle\tilde\varphi_n, (\exp 2f_n)
\eta_n\half[p^2,\eta_n]\tilde\varphi_n\rangle.
$$
Consequently,
$$
|\tilde \varphi_n(a,\kappa)|^2\leq (\exp-2f_n(a,\kappa))
\hbox{\rm Re}\langle\tilde\varphi_n, W_n\tilde\varphi_n\rangle,
$$
where
$$
W_n =(\exp 2f_n)\ \eta_n \half [p^2, \eta_n]= 
-\half (\exp 2f_n)(\eta_n \eta_n'' +2i \eta_n \eta_n' p),
$$
so that
$$
\hbox{\rm Re} W_n =
-\half \exp 2f_n(\eta_n \eta_n'' - 2f_n' \eta_n\eta_n' - (\eta_n \eta_n')').
$$
It follows from the support properties of $\eta_n'$ and the definition of $f_n$ 
that
there exists a constant $C_n$ so that for all $\kappa$, one has $|\hbox{\rm Re} 
W_n|\leq C_n$. As a result, a simple computation shows that, for all 
$\epsilon$, 
there exists  constants $C_{n,\epsilon}, K_{n,\epsilon}$ so that for all 
$\kappa>K_{n,\epsilon}$, and for all $0\leq \tilde x < x_n(\kappa)-2$
$$
|\tilde\varphi_n(\tilde x,\kappa)|^2\leq C_{n,\epsilon} \exp\left \{ - 
\half(1-\epsilon)(\kappa-\tilde x)^2\right\}.
$$
This proves the first statement of (iv). To prove the second estimate, it is
clear from part (ii) that we need to prove the above estimate holds for 
$|\tilde\varphi_n'(\tilde x,\kappa)|^2$ as well. If $\chi$ is a smooth 
characteristic function of the interval $[0,1]$ with support in $[0,2]$, one 
has, using the eigenvalue equation and two partial integrations that
\begin{eqnarray*}
 |\tilde\varphi_n'(0,\kappa)|^2&\leq& \int_0^1|\tilde\varphi_n'(\tilde 
x,\kappa)|^2 d\tilde x \\
&\leq& \int_0^\infty \chi(\tilde x)  
\tilde\varphi_n'(\tilde x,\kappa)^2\leq C \int_0^1|\tilde\varphi_n(\tilde 
x,\kappa)|^2 d\tilde x,
\end{eqnarray*}
from which the result follows. The last part of the Lemma is an immediate 
consequence of the Feynman-Hellman formula.
\qed

\noindent {\bf Acknowledgements:} The authors would like to thank Forbairt,
the Royal Irish Academy, the CNRS and the Minist\`ere des Affaires Etrang\`eres 
for their financial support. They also thank J.Fr\"ohlich, G.M.Graf and
J.Walcher for bringing their results to their attention prior to publication.
\newpage

\end{document}